# Extended Statistical Thermal Model and Rapidity Spectra of Hadrons at 200 GeV/A

Saeed Uddin[!], Majhar Ali, Jan Shabir[t]

*Department of Physics, Jamia Millia Islamia, New Delhi-110025*

M. Farooq Mir

*Department of Physics, University of Kashmir, Srinagar, J & K*

## ABSTARCT

We use the extended statistical thermal model to describe various hadron's rapidity spectra at the highest RHIC energy (200 GeV/A). The model assumes the formation of hot and dense regions moving along the beam axis with increasing rapidities, $y_{FB}$. It has been earlier shown that this model can explain the net proton flow i.e. $p - \bar{p}$, ratio $\bar{p}/p$ and the pion rapidity spectra. In this paper we have attempted to show that in addition to these quantities, this model can also successfully describe the individual rapidity spectra of protons, antiprotons, Kaons, antiKaons, pions, the ratios $\bar{\Lambda}/\Lambda$ and $\bar{\Xi}/\Xi$. The experimental data set on p, $\bar{p}$, K, $\bar{K}$ and $\pi$ provided by BRAHMS collaboration at the highest energy of Relativistic Heavy Ion Collider, $\sqrt{S_{NN}}$ = 200 GeV are used. The theoretical results fit quite well with mid-rapidity data (for |y| < 1) of the $\bar{\Lambda}/\Lambda$ and the $\bar{\Xi}/\Xi$ ratios available (from STAR). We have used single set of model parameters including single value of the temperature parameter T for all the regions of the hot and dense matter formed. The chemical potentials are however assumed to be dependent on the fireball rapidity $y_{FB}$. We have analyzed the contribution of the decay of the heavier resonances to the proton (antiproton) rapidity spectra. It is found that the rapidity spectrum of the product hadron is nearly same as that of the parent hadron. We have also imposed the criteria of exact strangeness conservation in every (local) region of the dense matter separately, which is necessary. We also discuss what can be learned about the nuclear transparency effect at the highest RHIC energy from the net proton rapidity distribution.

[!] E-mail : saeed_jmi@yahoo.co.in

[†] On leave of deputation from Department of Physics, Amar Singh College, Srinagar, J & K, India.

# 1. INTRODUCTION

The yields of baryons and antibaryons are an important indicator of the multi-particle production phenomenon in the ultra-relativistic nucleus-nucleus collisions. Great amount of experimental data have been obtained in such experiments ranging from the AGS energies to the RHIC. The study of ultra-relativistic nuclear collisions allows us to learn how baryon numbers initially carried by the nucleons only (before the nuclear collision) are distributed in the final state [1] in various baryonic states at the thermo-chemical freeze-out after the collision. It is also possible to obtain important information about the energy loss of the colliding nuclei by analyzing the rapidity dependence of the p and $\bar{p}$ production. The study of *net* proton ($p - \bar{p}$) and other hadron's rapidity spectra in such experiments can throw light on the collision scenario, namely the extent of nuclear stopping/transparency, the subsequent formation of the hot and dense regions of hadronic matter (or fireballs) and the degree of thermo-chemical equilibration of various hadronic resonances. The degree of chemical equilibration of hadronic resonances can also be estimated by their contributions to the lighter hadrons viz. $p, \bar{p}, K$ etc. via their decay. The experimental data on the net proton flow at the AGS energy showed a peak at midrapidity, while at the top SPS energy ($\sqrt{S_{NN}} \approx 17.2$ GeV) the distribution started showing a minimum at midrapidity. The SPS data at different energies (20, 30, 40, 80, 158 GeV/A) show [1] that at midrapidity the $p - \bar{p}$ yield decreases gradually with increasing energy. This implies that at SPS energies the nuclear collisions start exhibiting some transparency. Hence this new property namely the extended longitudinal scaling in the rapidity distributions has emerged [2 - 5]. This has been observed in pp collisions at the highest RHIC energies as well as ultra-relativistic nuclear collisions [5, 6]. Data from the BRAHMS collaboration [7] clearly show that the antiproton to proton ratio shows a maximum at midrapidity and gradually decreases towards larger rapidities, whereas the net proton flow shows a broad minimum, spanning approximately ± 1 unit around the midrapidity region of dN/dy spectra. It was therefore conjectured [7] that at RHIC energies the collisions are (at least partially) transparent. Though the midrapidity region at RHIC is not yet totally baryon free however a transition from a baryon dominated system at lower energies to an *almost* baryon free system in the midrapidity at RHIC can be observed. An interesting analysis by Stiles and Murray [2, 8] shows that the data obtained by the BRAHMS collaboration at 200 GeV/A has a clear dependence of the baryon chemical potential on rapidity which is revealed through the changing $\bar{p}/p$ ratio with rapidity. Biedroń and Broniowski [2,9] have also done an analysis of rapidity dependence of the $\bar{p}/p$, $K^+/K^-$, $\pi^+/\pi^-$ ratios based on a single freeze-out model of relativistic nuclear collisions. They have used a single fireball model where the baryon chemical potential depends on the spatial rapidity $\alpha_{||}$ = arctanh(z/t) inside the fireball. They use single temperature parameter for the entire fireball at the freeze-out (≈ 165 MeV).

Their model is very successful in explaining the proton, antiproton and the (anti)Kaon rapidity spectra. They have used a parameterization for the functional dependence of the chemical potentials at low values of $|\alpha_{||}|$ as $\mu = \mu(0)[1 + A\, \alpha_{||}^{2.4}]$. Here the power is chosen to be 2.4 instead of 2 which according to these authors works better. Fu-Hu Liu *et al.* have [10] recently attempted to describe the transverse momentum spectrum and rapidity distribution of net protons produced in such high energy nuclear collisions by using a new approach viz. a two cylinder model [11 - 13].

## 2. MODEL

We briefly discuss the model used here. In order to describe the rapidity distribution of the produced hadrons in ultra-relativistic nuclear collisions the statistical thermal model has been extended to allow for the chemical potential and temperature to become *rapidity dependent* [5, 9, 14 - 16]. The model assumes that the rapidity axis is populated with hot regions (fireballs) moving along the beam axis with increasing rapidity, $y_{FB}$. The emitted particles leave these regions (fireballs) at the freeze-out following a (local) thermal distribution. The resulting rapidity distribution of any given particle specie (say $j^{th}$) is then obtained by a superposition of the contributions of these regions (fireballs) as follows :

$$\frac{dN^j(y)}{dy} = A \int_{-\infty}^{+\infty} \rho(y_{FB}) \frac{dN_1^j(y-y_{FB})}{dy} dy_{FB} \qquad (1)$$

where y is the particle's rapidity in the *rest frame* of the colliding nuclei and A is the overall normalization factor. The distribution $\frac{dN^j(y)}{dy}$ represents total contribution of all the regions to the $j^{th}$ hadron specie's rapidity spectra. According to the thermal model the single particle rapidity spectra of the hadrons in these local regions (fireballs) i.e. $\frac{dN_1^j}{dy}$ can be written as :

$$\frac{dN_1^j}{dy} = 2\pi g^j \lambda^j \left[ m_o^2 T + \frac{2m_o T^2}{Coshy} + \frac{2T^3}{Cosh^2 y} \right] e^{-\beta m_o Coshy} \qquad (2)$$

where T is the temperature of the fireball. The $m_o$, $g^j$ and $\lambda^j$ are the rest mass, spin-isospin degeneracy factor and the fugacity of the $j^{th}$ hadronic specie, respectively.

Here the contribution of the respective regions to the final hadronic yields is not assumed to be in the equal proportions. It is rather assumed to follow a Gaussian distribution in the variable $y_{FB}$, centred at zero fireball rapidity ($y_{FB}$ = 0) :

$$\rho(y_{FB}) = \frac{1}{\sqrt{2\pi}\sigma} \exp\left( \frac{-y_{FB}^2}{2\sigma^2} \right) \qquad (3)$$

The value of σ determines the width of the Gaussian distribution. Furthermore the experimental data provide a strong evidence that the baryon chemical potential ($\mu_B$) of the successive regions (fireballs) should be dependent on their rapidities ($y_{FB}$). Hence the evaluation of the final hadronic yield requires a superposition of the contributions of all these regions (fireballs) whose baryon chemical potential increases with their rapidity. For this purpose a quadratic type dependence is considered [2, 4], which also leaves the chemical potential ($\mu_B$) invariant under the transformation $y_{FB} \rightarrow -y_{FB}$ :

$$\mu_B = a + b\, y_{FB}^2 \tag{4}$$

In the recent works [5, 18] it has been further assumed that the temperature of the successive fireballs along the rapidity axis *decreases* (as the baryon chemical potential increases) according to a chosen parameterization of the following type :

$$T = 0.166 - 0.139\, \mu_B^2 - 0.053\, \mu_B^4 \tag{5}$$

where the units are in GeV. Here the temperature of the mid-rapidity region of the fireball with $y_{FB} \sim 0$ is fixed at $T = (0.166 + 0.139\, a^2 - 0.053\, a^4)$ GeV. Becattini and Cleymans [2, 4] provided a good fit to the net proton flow, pion flow and the ratio $\bar{p}/p$, measured at the highest RHIC energy by the BRAHMS collaboration. The values of their model parameters were, a = 23.8 MeV, b =

11.2 MeV and $\sigma$ = 2.183, while the temperature T varies according to the above parameterization (Eqn. 5). However a theoretical fit to the individual proton and antiproton rapidity spectra was not provided. Furthermore the strange sector data on Kaon and antiKaon rapidity spectra as measured by the BRAHMS collaboration in the same experiment under the same experimental conditions and the ratios $\bar{\Lambda}/\Lambda$ and $\bar{\Xi}/\Xi$ (from STAR) were also not tested.

Therefore in this letter we have attempted to show that in addition to the net proton flow, pion flow and the ratio $\bar{p}/p$ the extended statistical thermal model discussed above can also very effectively explain the *individual* rapidity spectra of various non-strange and strange hadrons such as the protons, antiprotons, Kaons, antiKaons, $\bar{\Lambda}/\Lambda$, and the $\bar{\Xi}/\Xi$. We find that the mid-rapidity data (for |y| < 1) available (from STAR) on $\bar{\Lambda}/\Lambda$ and $\bar{\Xi}/\Xi$ are fitted quite well. This can be achieved with single set of model parameters viz. a, b, $\sigma$ and single value of the temperature parameter T chosen for all the regions of the dense hadronic matter (fireballs). The chemical potentials have to be made dependent on the fireball rapidity $y_{FB}$, a situation which is unavoidable in the model due to the nature of the experimental data.

We have also investigated the contribution of the decay products of the heavy resonances [17] to the proton and antiproton rapidity distributions. The BRAHMS experimental data on the proton and antiproton also include the contribution of the weak decays, mainly from the $\Lambda$.

The spectrum of a decay product of a given parent hadron in the *rest frame of the fireball* can be written as [17] :

$$\frac{d^3 n^{decay}}{d^3 p} = \frac{1}{2pE}\left(\frac{m_h}{p^*}\right)\int_{E_-}^{E_+} dE_h \, E_h \left(\frac{d^3 n_h}{d^3 p_h}\right) \qquad (6)$$

where *p* and *E* are respectively the momentum and energy of the product hadron in the rest frame of the fireball with $p=(E^2-m^2)^{1/2}$. The subscript *h* in Eqn. (6) stands for the decaying (parent) hadron. The two body decay kinematics gives the **product** hadron's momentum ($p^*$) and energy ($E^*$) in the *rest frame of the decaying hadron* as:

$$p^* = (E^{*2}-m^2)^{1/2}, \qquad (7)$$

$$E^* = \frac{m_h^2 - m_k^2 + m^2}{2m_k}, \qquad (8)$$

where $m_k$ indicates the mass of the **other** decay hadron produced along with the one under consideration (having mass *m*). Thus the limits of integration in the Eqn. (6) are :

$$E_\pm = \left(\frac{m_h}{m^2}\right)\left(EE^* \pm pp^*\right), \qquad (9)$$

Note again that $E$ ($E_h$) and $p$ ($p_h$) are, respectively, the product hadron's (decaying hadron's) energy and momentum in the "rest frame of the *fireball*". In our analysis we have also applied the criteria of exact strangeness conservation in the dense hadronic matter. It is done in a way such that the net strangeness is zero not only on the overall basis but also in every region (fireball) separately. This is essential because as the rapidity $y_{FB}$ of these regions formed increases along the rapidity axis, the baryon chemical potential ($\mu_B$) also increases. Hence the required value of the strange chemical potential ($\mu_S$) varies accordingly with $\mu_B$ for a given value of temperature. Consequently the value of the strange chemical potential ($\mu_S$) varies with $y_{FB}$.

## 3. RESULTS AND DISCUSSION

In figures 1 and 2 we have shown the experimental dN/dy data by the red solid circles for the protons and antiprotons, respectively, obtained from the top 5% most central collisions at $\sqrt{S_{NN}}$ = 200 GeV in the BRAHMS experiment. The errors are both statistical and systematic. The experimental data are symmetrized for the negative values of rapidity [7]. The proton's and antiproton's *dN/dy* data show a maximum at midrapidity and decrease towards higher rapidities (y ~ 3). The proton (antiproton) experimental data also include the contributions of the weakly decaying hyperons. Using the MC simulations BRAHMS have deduced that there are 0.53 ± 0.05 protons for each Λ and 0.49 ± 0.05 protons for each $\Sigma^+$ decay. Here lambdas also include the contributions from other hyperons that decay to protons through their decays to Λs, i.e. $\Sigma^0$, $\Xi^0$ and $\Xi^-$ [7]. The branching ratio of the decay of these particles into lambda is ≈ 100%. We have considered all hadrons up to the mass 1530 MeV in full chemical equilibrium.

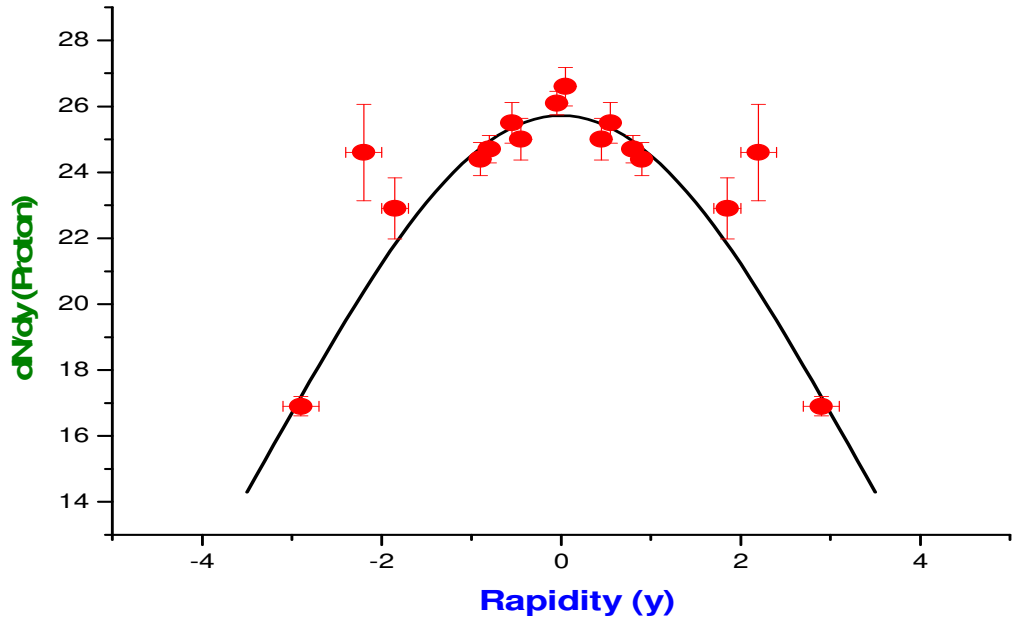

*Fig. 1 : Proton rapidity spectra. Theoretical result is shown by the solid curve while the red solid circles represent the experimental data points.*

We have fitted both the spectral shapes in figures 1 and 2 simultaneously for, *a* = 25.5, *b* = 12.7, *σ* = 2.1 and T = 175.0 MeV.

We find that the theoretical curves fit the data quite well in both the cases. The theoretical curves also include the contribution of the resonance decays in accordance with the BRAHMS suggestions, discussed above.

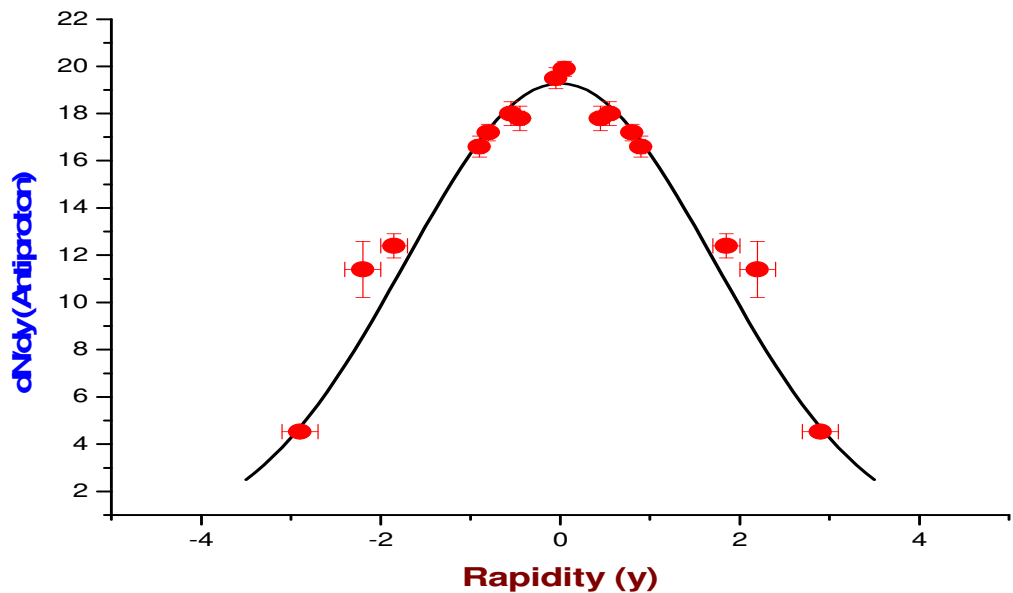

*Fig. 2 : Antiproton rapidity spectra. Theoretical result is shown by the solid curve while the red solid circles represent the experimental data points.*

We have used the same normalization factor (Eqn. 1) A = 38.5 for all the baryons and antibaryons. This also minimizes the value of (weighted) $X^2$/DoF of the theoretical curves in figures 1 and 2. The minimum (weighted) $X^2$/DoF for the fitted curves in figures 1 and 2 are 3.04 and 4.81, respectively. The somewhat larger value for the case of antiproton is due to the small error bars.

It may be noted that the proton spectra (experimental as well as theoretical) in the figure 1 is seen to be slightly broader than the antiproton spectra in figure 2. This according to the present model seems to emerge from the fact that since $\mu_B \sim y_{FB}^2$ and the rapidity axis is assumed to be populated by the regions (fireballs) of successively increasing rapidity $y_{FB}$ and hence increasing chemical potentials (since $\mu_B = a + b\, y_{FB}^2$), consequently the low rapidity (y) baryons (which have a larger population in a baryon rich fireball in thermo-chemical equilibrium) emitted in the forward (backward) direction from a *fast* moving region (fireballs with large $y_{FB}$), appear with a large value of rapidity (y) in the *rest frame of the colliding nuclei*. In other words as the baryon chemical potentials ($\mu_B$) increase monotonically along the rapidity axis ($\sim y_{FB}^2$) there is an increase in the *density* of the protons and a simultaneous suppression in the *density* of antiprotons thereby making the proton rapidity spectrum broader than that of the antiproton's.

We also compare these theoretical spectral shapes of the dN/dy distributions of protons and antiprotons with the cases when resonance decay contributions are not taken into account. In figures 3 and 4 we have shown the theoretical rapidity distributions of protons and antiprotons of *pure thermal* [17] origins, respectively. The *pure thermal* yield is the one which does not include the resonance decay contributions, as discussed above. This is shown by the red curves. We also show again the *total* proton and antiproton distributions by green curves (i.e. including the resonance decays, as shown in figures 1 and 2 also).

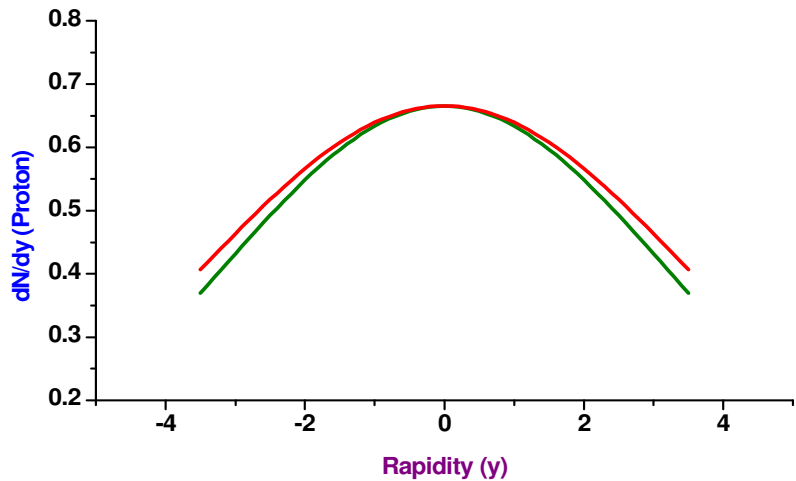

*Fig. 3 : Rapidity spectra of pure thermal protons (i.e. excluding the resonance decay contributions) is shown by red curve and that of the total protons is shown by green curve (i.e. including the resonance decays). For the sake of proper comparison of the spectral shapes the two curves are normalized at y = 0. The units are arbitrary here.*

To facilitate a proper comparison the curves have been normalized at y = 0 in both the figures, where the units are arbitrary. We notice that the spectral shapes for the case when resonance decay contributions are taken into account are (slightly) narrower than the spectral shape of the protons of the pure thermal origin, while for the case of antiprotons (figure 4) the two curves almost overlap. Hence we find no major change of rapidity spectral shape due to resonance contributions.

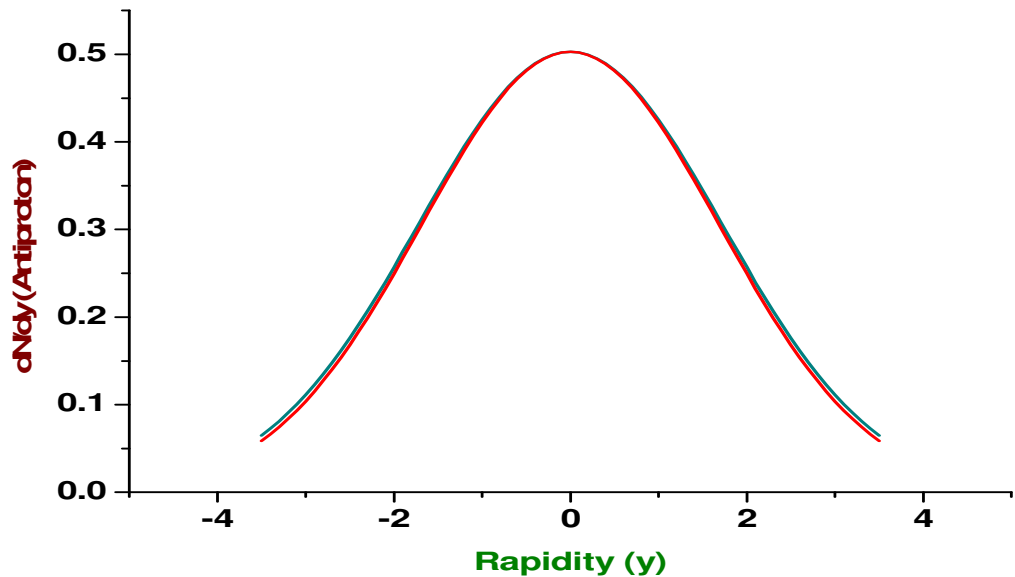

*Fig. 4 : Rapidity spectra of pure thermal antiprotons (i.e. excluding the resonance decay contributions) shown by red curve and that of the total antipro tons is shown by green curve (i.e. including the resonance decays). For the sake of proper comparison of the spectral shapes the two curves are normalized at y = 0. The units are arbitrary here.*

In order to highlight this further we have in figure 5 plotted the *ratio* of the two (i.e. *total* protons versus *pure thermal* protons) with the rapidity. We find that the ratio is nearly constant between y = 0 to y ≈ 3. The values represent the actual results of the calculations.

The variation in the ratio between these limits of the rapidity variable is < 6%. The corresponding ratio for the case of antiprotons will be almost constant.

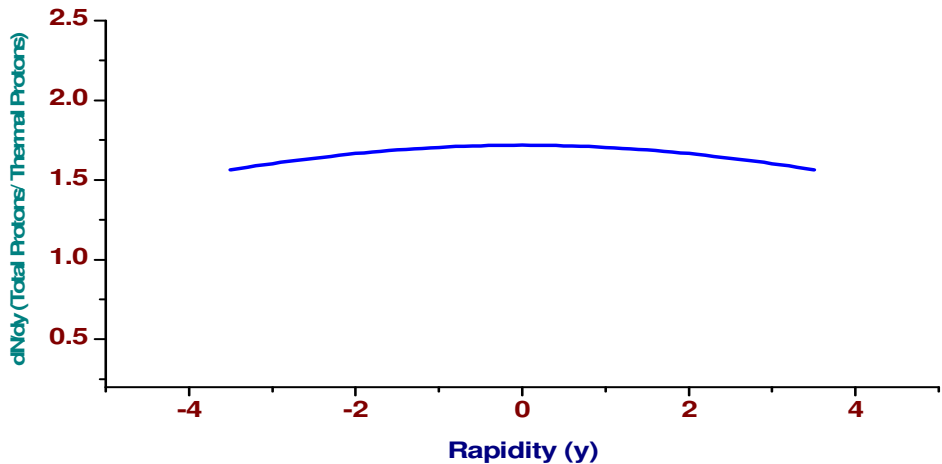

*Fig. 5 : Ratio of the total protons versus pure thermal protons plotted with rapidity. The values represent the actual results.*

We have found that nearly 66% protons (antiprotons) are of the *pure thermal* origin. This agrees with the experimental situation within the limits of statistical and systematic errors [7].

The slight narrowness in the spectral shape of the total protons is due to the relative narrowness in the rapidity distributions of the heavier baryons such as Λ, Σ, Ξ compared to the lighter particles i.e. protons.

To realize this we have shown in figure 6 the rapidity distributions of *pure thermal* p, Λ, Σ, and Ξ. We have normalized these curves at y = 0 for the sake of comparison of the spectral shapes.

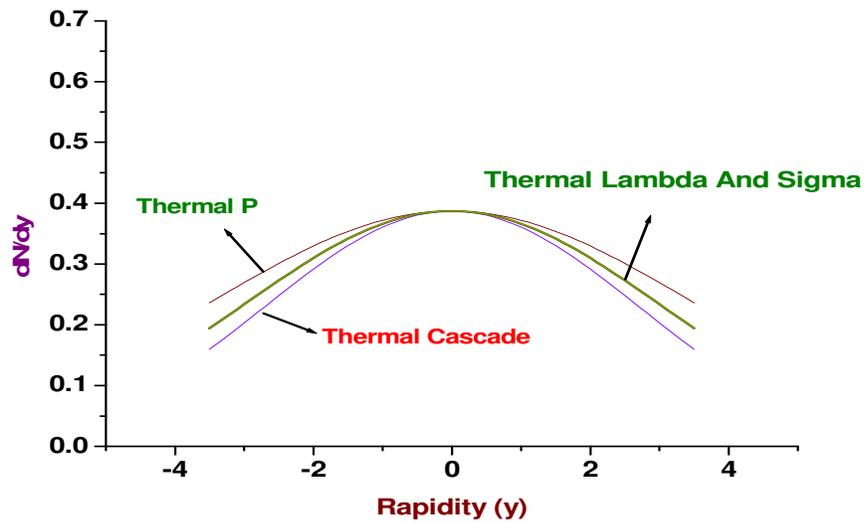

*Fig. 6 : Rapidity distributions of pure thermal p, Λ, Σ, and Ξ. For a proper comparison of the spectral shapes we have normalized these curves at y = 0.*

We see that the spectral shape of Ξ is narrower than that of the proton's and lambda's (sigma's) for the given values of the parameters a, b, σ and T, while the curves of lambdas and sigmas overlap, since their masses are not much different. In figure 7 we have plotted the ratio of pure thermal cascades and lambdas with rapidity. We find that the ratio does not vary significantly between y = 0 and y ≈ 3. The values are the actual result of the theoretical calculations.

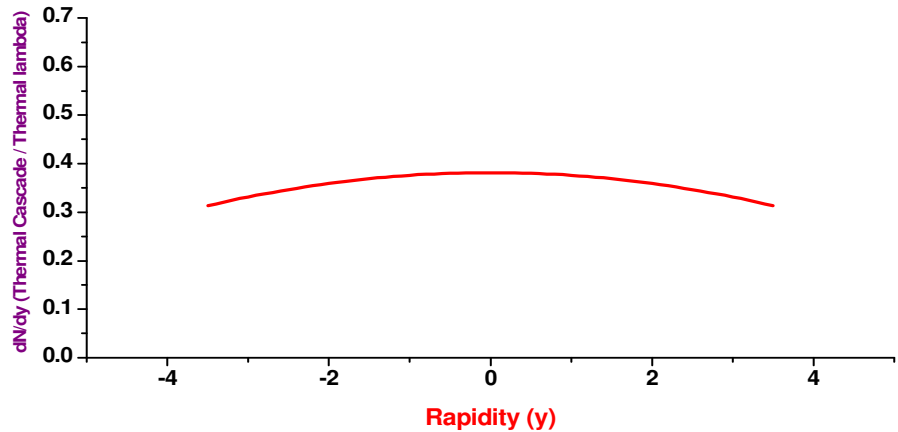

*Fig. 7 : Ratio of pure thermal Cascades and lambdas. The values are the result of actual calculations. The ratio varies slowly with y.*

Next we have shown in figure 8 the rapidity distribution shapes of the *pure thermal lambdas* in the extended statistical model and that of the *protons obtained through its decay*.

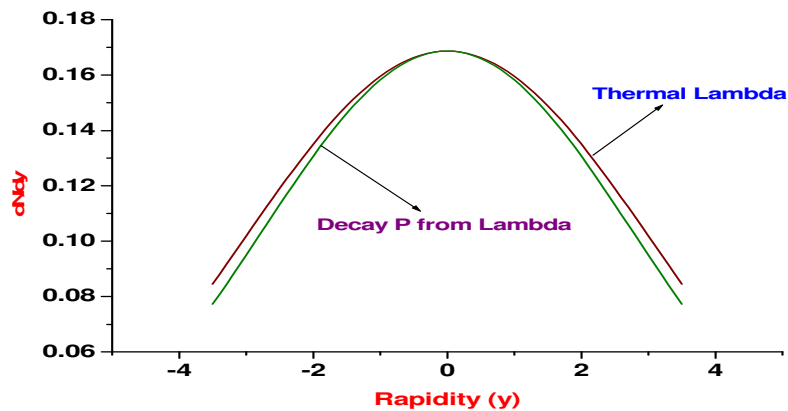

*Fig. 8a : Rapidity distribution shapes of the pure thermal lambdas and that of the protons obtained through its decay. The two curves are normalized at y = 0 for the sake of comparison of the spectral shapes.*

The two curves are though very much similar in shape however the decay protons have a slightly narrower spectrum than that of the parent particle's (i.e. the lambda). The two curves in figure 8a are again normalized to enable a proper comparison of the spectral shapes. In figure 8b we have shown the rapidity spectra of the *pure thermal Cascades* in the extended statistical model and that of the *lambdas obtained through its decay*. The two curves in figure 8b are normalized to enable a proper comparison of the spectral shapes.

We therefore conclude that the rapidity spectra of the product hadrons are only slightly narrower than those of the parent hadrons.

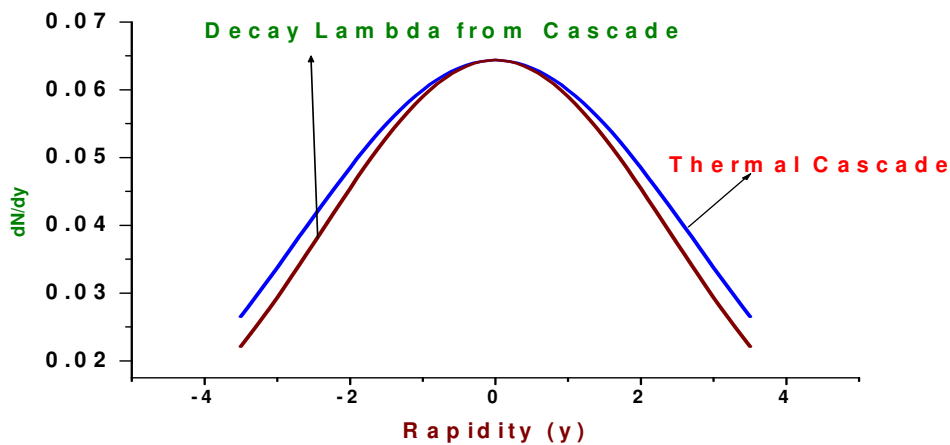

*Fig. 8b : Rapidity distribution shapes of the pure thermal cascades and that of the Lambdas obtained through its decay. The two curves are normalized at y = 0 for the sake of comparison of the spectral shapes. The units are arbitrary.*

In figure 9 we have shown the theoretical and the experimental results of the net proton flow. The solid curve represents the result of our calculation while the solid red circles represent the BRAHMS experimental data. The value of the minimum (weighted) $\chi^2$/DoF for the fitted curve is 3.12. We have again used the same normalization factor i.e. 38.51 for all baryons and antibaryons. We find that in our theoretical calculation the net baryon number $N_B \approx 2.18\, N_p$, where $N_p$ is the theoretically calculated *net* proton number. This is in close agreement with the BRAHMS estimate where they have suggested $N_B \approx (2.03 \pm 0.08)\, N_{p,meas}$, where $N_{p,meas}$ is the experimentally measured net proton number. The experimental net proton flow distribution data shows a somewhat broad and deep minimum around the mid-rapidity region which is well described by the theoretical solid curve in figure 9. This experimental situation is however not as it was widly expected [19] that the rapidity distribution of the *net baryons* produced in the ultra-relativistic nuclear collisions will exhibit a very flat and broad minimum measuring several units of rapidity, centred at midrapidity. So far it has not been observed, either in the SPS experiments or even at the highest RHIC energy. The future LHC/ALICE experiments may throw more light on this aspect and provide a better understanding of the nuclear transparency effect in the ultra-relativistic nuclear collisions.

The present RHIC experiments at 200 GeV/A however have given an indication that these nuclear collisions have begun to show *at least partial transparency*.

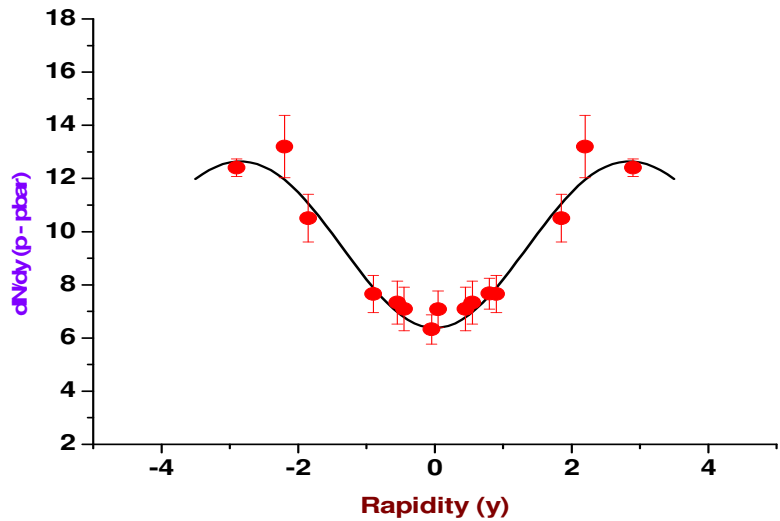

*Fig. 9 : Rapidity spectra of the net proton flow. The solid red circles represent the BRAHMS experimental data points while the solid curve shows the result of theoretical calculation.*

In figure 10 we have shown by the solid curve the theoretical rapidity spectra of $\bar{p}/p$ ratio. The solid green circles are the experimental data points. The ratio has a somewhat broad maximum (~ 0.75) in the midrapidity region which then decreases to about 25% at around y ~ 3. We have included the resonance decay contributions. The value of the minimum (weighted) $X^2$/DoF for the curve is 1.06.

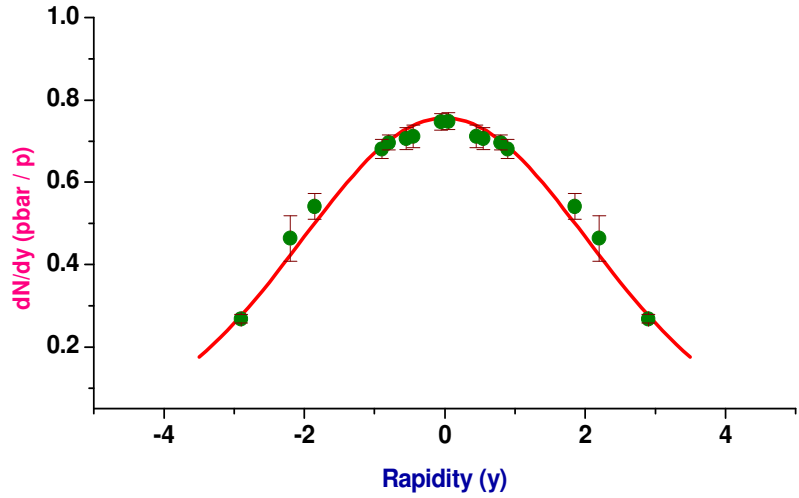

*Fig. 10 : Rapidity spectra of $\bar{p}/p$ ratio. The solid curve is the result of theoretical calculation. The solid green circles represent the BRAHMS experimental data points.*

Apart from the analysis of the rapidity spectra of non-strange baryons we have also analyzed the strange meson data as measured by the BRAHMS collaboration at RHIC in the top 5% most central collisions at 200 GeV/A in the same Au - Au collision experiments. We have found that in the extended statistical thermal model it is possible to account fully for the rapidity distributions of Kaons and antiKaons also.

In figure 11 we have shown the rapidity spectra of Kaon flow. The theoretical curve which fits the data is for the *same* values of the model parameters as used for the theoretical curves in all previous figures 1-10. We find that the theoretical curve provides a very good fit to the data. Here it is noteworthy that the BRAHMS experimental data of Kaons and antiKaons used here have been *corrected* for the decay contributions. Hence in our calculations we have not included the contribution of the strong $K^* \to K\pi$ decay. The $X^2$/DoF for the fitted curve is 0.70.

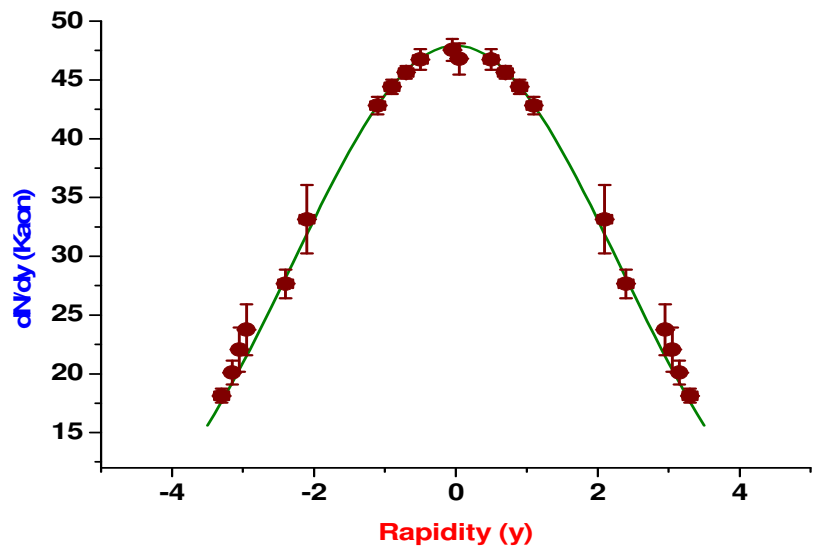

*Fig. 11 : Rapidity spectra of Kaon flow. The theoretical curve which fits the data is for the same values of the model parameters as used for the theoretical curves in figures 1 - 10.*

In figure 12 we have shown the rapidity spectra of antiKaons. The theoretical curve which fits the data is again for the same values of the model parameters. Here once again we notice that the Kaon spectra is broader than the antiKaon spectra as is also noticed for the proton spectra which is broader than the antiproton spectra. This is again due to the increasing chemical potential of the successive regions (fireballs) formed along the rapidity axis. The $X^2$/DoF for the fitted curve is 4.69. The normalization factor obtained for Kaons and antiKaons is ≈ 45.6.

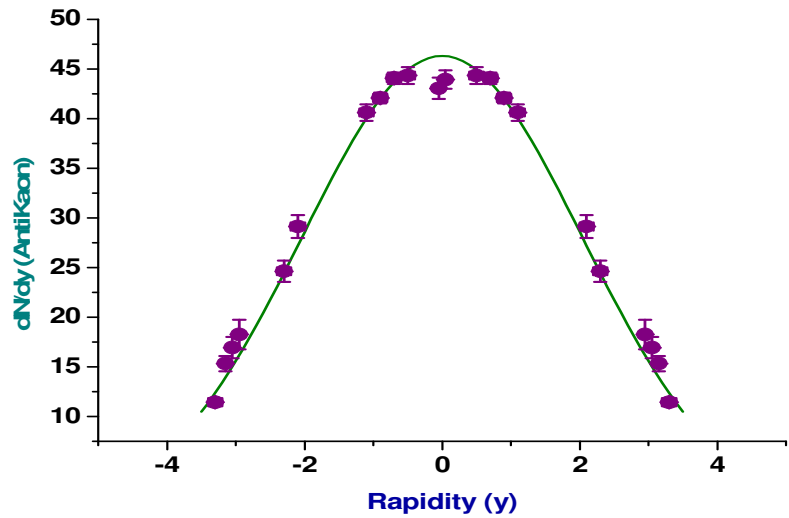

*Fig. 12 : Rapidity spectra of antiKaon flow. The theoretical curve which fits the data is for the same values of the model parameters as used for the theoretical curves in all previous cases.*

In figure 13 we have shown the theoretical rapidity spectra of the ratio $\bar{\Lambda}/\Lambda$, which includes the decay contributions of $\Sigma$ and $\Xi$ resonances.

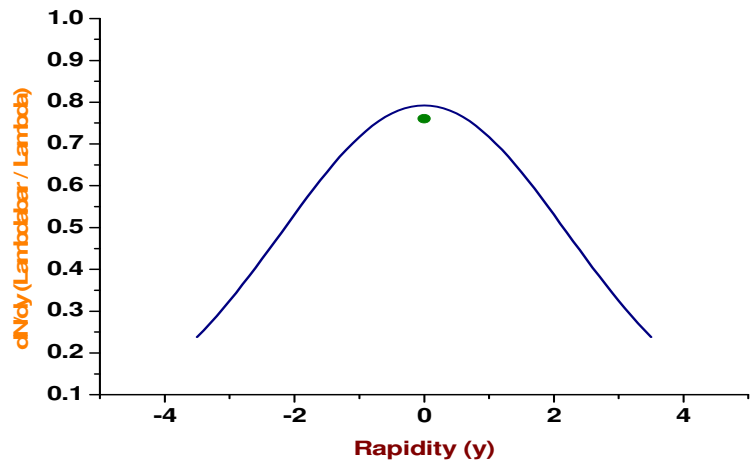

*Fig. 13 : Rapidity spectra of $\bar{\Lambda}/\Lambda$ flow. The midrapidity data point from the STAR is shown.*

The Fig.14 shows the theoretical rapidity spectra of the ratio $\bar{\Xi}/\Xi$. The mid-rapidity data (for |y| < 1) available from STAR [20] fit quite well in both these

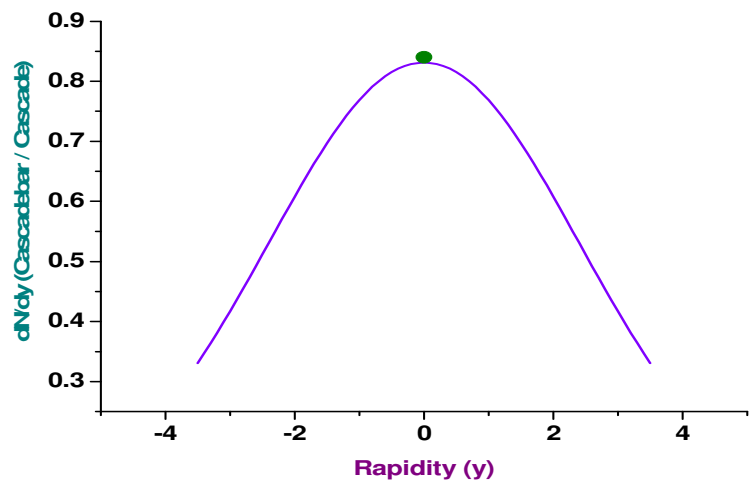

*Fig. 14 : Rapidity spectra of $\bar{\Xi}/\Xi$ flow. The midrapidity data point from the STAR is also shown.*

cases. The theoretical curves are for the same values of the model parameters as the ones used for protons, antiprotons, Kaons and antiKaons.

In figure 15 we have shown the theoretical fit to the pion's experimental data. The pion spectrum is also fitted for the same values of the model parameters as used in other cases. Here it is again important to note that the BRAHMS have corrected these pion experimental data also for any feed down contributions from heavier resonances including decaying baryons like Λ, Σ, Ξ. Hence this facilitates a direct comparison of the pion experimental data with the calculated abundance of the pions of the *pure thermal* origin (i.e. excluding the decay contributions). It should be realized that under such a situation the pion rapidity spectrum is then sensitive only to the two model parameters, which are T and σ. The values of "*a*" and "*b*" have no role to play whatsoever in determining the pion's spectra here, since the parameters "**a**" and "**b**" are used to fix the baryon chemical potential in the different regions (fireballs) and the experimental data does not include the decay contributions of baryons and heavier mesons (K* etc). Consequently these parameters do not govern the abundance of the pions of the *pure thermal* origin. The $X^2$/DoF for the fitted curve is 6.5, which is somewhat high due to the small error bars in the experimental data. The normalization factor A for this case turns out to be ≈ 97.5.

We notice that in the present model the normalization factors are not same for all the particle species. We find that for baryons it is ≈ 38.5, for (anti)Kaons it is ≈ 45.6 while for pions it turns out to be ≈ 97.5. However, in the present analysis the actual emphasis is on the capability of the model to reproduce the shape of the rapidity distributions, and this is fitted here for each particle specie.

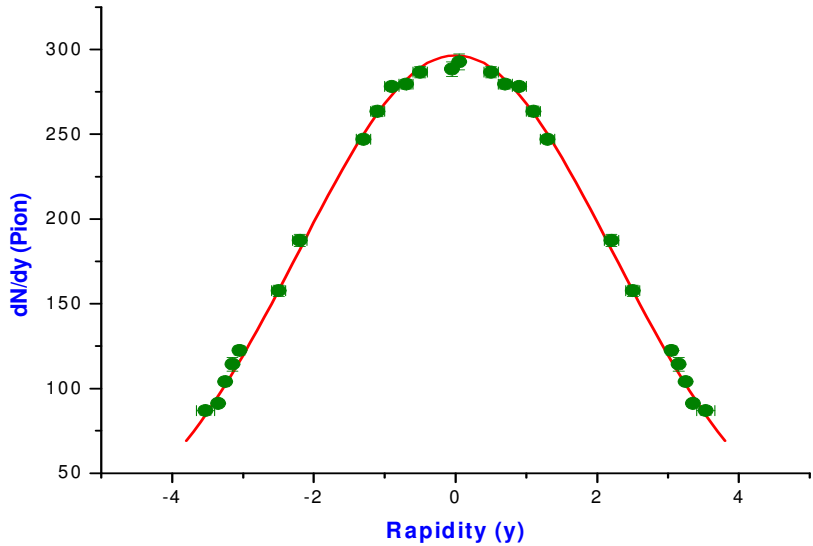

*Fig. 15 : Rapidity spectra of pion flow. The theoretical result is shown by the solid curve.*

Finally in figure 16 we have shown the variation of the $\mu_S$ with $y_{FB}$. The value of $\mu_S$ is determined by applying the strangeness conservation criteria for a given value of $y_{FB}$, which fixes $\mu_B$ since $\mu_B = a + b\, y_{FB}^2$. It is seen to first rise smoothly with $y_{FB}$ reaching a maximum value of about 30 MeV at around $y_{FB} \sim 5$ and then drops rapidly to very small values as $y_{FB} \sim 8$ (i.e. when $\mu_B \approx 800$ MeV) and beyond this will become negative.

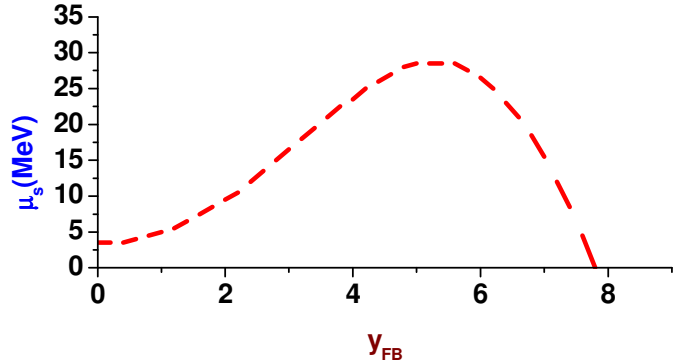

*Fig. 16 : The variation of the $\mu_S$ with $y_{FB}$ for T = 175.0 MeV.*

In summary, we use an earlier proposed extended thermal model, where formation of several hot regions (or fireballs) moving with increasing rapidity ($y_{FB}$) along the beam axis is assumed. The final state hadrons are assumed to be emitted from these fireball regions. A Gaussian profile in $y_{FB}$ is used to weigh the contributions of these regions to the final state emitted hadron's population. A quadratic profile in $y_{FB}$ is used to fix the baryon chemical potentials of these regions (fireballs). We find that it is possible to explain not only the net proton, $\bar{p}/p$ and pion flow but also the individual proton, antiproton, Kaon, antiKaon, $\bar{\Lambda}/\Lambda$ and the $\bar{\Xi}/\Xi$ rapidity spectra as well. It is interesting to find that the model can successfully explain the strange sector data also quite well, measured in the same experiment by the BRAHMS and the STAR collaboration. This is achieved by using *single* set of the model parameters. We also study the effect of the resonance decay products on the rapidity spectra of the hadrons. We find that the rapidity spectra of the decay products are slightly narrower than that of the parent hadron's. The pure thermal rapidity distribution of the heavier particle species is seen to be narrower than those of the lighter paticles.

*Authors are grateful to Professors J. Cleymans and F. Becattini for making some of the experimental data available to us. Jan Shabir is grateful to University Grants Commission, New Delhi, for the financial assistance during the period of deputation. Majhar Ali is thankful to University Grants Commission for providing Project Fellowship. Saeed Uddin is thankful to the University Grants Commission (UGC), New Delhi, for the Major Research Project grant.*